\documentclass[aps,prd,superscriptaddress,11pt,showpacs,notitlepage]{revtex4}
	\usepackage{graphicx}
	\usepackage{amsmath,amssymb,mathrsfs}
	\usepackage[T1]{fontenc}
\usepackage{epsf}
\usepackage{epsfig}
\usepackage{xcolor}
	\usepackage[colorlinks=true, a4paper=true, pdfstartview=FitV,
linkcolor=blue, citecolor=blue, urlcolor=blue]{hyperref}

\numberwithin{equation}{section}

\newcommand{\be}{\begin{equation}}
\newcommand{\ee}{\end{equation}}
\newcommand{\bea}{\begin{eqnarray}}
\newcommand{\eea}{\end{eqnarray}}

\newcommand{\bb}{\bibitem}
 
\newcommand{\eqn}{\begin{eqnarray}}
\newcommand{\eqnx}{\end{eqnarray}}

\begin{document}
\title{Gauged BPS baby Skyrmions with quantised magnetic flux}
\author{C. Adam}
\affiliation{Departamento de F\'isica de Part\'iculas, Universidad de Santiago de Compostela and Instituto Galego de F\'isica de Altas Enerxias (IGFAE) E-15782 Santiago de Compostela, Spain}
\author{A. Wereszczynski}
\affiliation{Institute of Physics,  Jagiellonian University,
Lojasiewicza 11, Krak\'{o}w, Poland}

\begin{abstract}
A new type of gauged BPS baby Skyrme model is presented, where the derivative term is just the Schroers current (i.e., gauge invariant and conserved version of the topological current) squared. This class of models has a topological bound saturated for solutions of the pertinent Bogomolnyi equations supplemented by a so-called superpotential equation. In contrast to the gauged BPS baby Skyrme models considered previously, the superpotential equation is linear and, hence, completely solvable. Furthermore, the magnetic flux is quantized in units of $2\pi$, which allows, in principle, to define this theory on a compact manifold without boundary, unlike all gauged baby Skyrme models considered so far. 
\end{abstract}
\maketitle 

\section{Introduction}
The baby Skyrme model \cite{baby} (for recent developments see \cite{babySk}) is a lower-dimensional version of the Skyrme model \cite{skyrme} which allows to study certain properties of the latter theory in a simplified environment. The Skyrme model, in turn, is a candidate low energy effective model of QCD which provides an efficient and very predictive framework for the study of properties of baryons \cite{baryon}, as well as atomic nuclei \cite{nuclei}, nuclear matter \cite{matter} and neutron stars \cite{stars}. Owing to its simplicity, the baby Skyrme model can be considered a laboratory where many ideas about $3+1$-dimensional Skyrmions can be developed and tested. Furthermore, the baby Skyrme model is of some relevance in condensed matter physics, so its analysis is of some independent interest, beyond its use as a toy model. 

In its application to nuclear physics,
one of the most important challenges for the Skyrme model is its ability to reproduce the (small) physical values of the binding energies of atomic nuclei. 
One way to accomplish this aim is to move the model towards a BPS regime. This is possible, because the full (generalised) Skyrme model action contains a submodel which is a BPS theory, i.e., has solitonic solutions with energies linear in the topological charge (baryon number) \cite{skyrme BPS}. The classical binding energies are, therefore, zero by construction in the BPS submodel. Physical binding energies can then be obtained by the inclusion of a (small) contribution from the remaining terms of the full Skyrme model and/or by the inclusion of semiclassical contributions \cite{nearBPS}. Here, also the choice of the potential (non-derivative term) plays an important role, because certain
"repulsive" potentials may, by themselves, further reduce the binding energies \cite{loosely}. 
That is to say, the above "near-BPS" Skyrme model proposal corresponds to assuming that the BPS part of the Skyrme model provides the main contributions to the nuclear masses. Independently, it has also been shown that the BPS submodel governs the thermodynamics of Skyrmions in the high density/pressure regime \cite{term}. 

In line with its role as a toy model for Skyrmions, also
the baby Skyrme model has a BPS submodel, known as the BPS baby Skyrme model \cite{baby1} (see also \cite{baby2}, \cite{bMartin}). Its lagrangian reads (for  the moment we omit all coupling constants)
\be
\mathcal{L}_{baby \; BPS}= -\frac{1}{4} \left(  \epsilon_{\mu \nu \rho} \left( \vec{\phi} \cdot \partial^\nu \vec{\phi} \times \partial^\rho \vec{\phi} \right)   \right)^2 - V(\vec{n} \cdot \vec{\phi}) 
\ee
where the derivative part is the topological current squared
\be
j_\mu^{top}= \frac{1}{2}  \epsilon_{\mu \nu \rho} \left( \vec{\phi} \cdot \partial^\nu \vec{\phi} \times \partial^\rho \vec{\phi} \right)  ,
\ee
 $\vec\phi \in \mathbb{S}^2$ is a three component unit vector field, and $\vec{n}$ is a constant vector, here chosen as $\vec{n}=(0,0,1)$. $V$ is a potential which breaks the original $O(3)$ symmetry of the derivative terms down to $O(2)$ and has a discrete number of isolated vacua (in practice, we consider only one- and two-vacuum potentials). In comparison to the full baby Skyrme model, the BPS submodel does not contain the usual kinetic term, quadratic in derivatives, i.e., the $O(3)$ sigma model term,
\be
\mathcal{L}_{baby Skyrme} = \mathcal{L}_{O(3)} + \mathcal{L}_{baby BPS}, \;\;\;\; \mathcal{L}_{O(3)}= \frac{1}{2} (\partial_\mu \vec{\phi})^2 .
\ee
Static solutions of the BPS baby Skyrme model obey a Bogomolny equation and saturate a corresponding topological bound, leading to a linear relation between energies and topological charges \cite{baby1}, \cite{bMartin}, \cite{stepien}. Another interesting property of this model is that it describes a perfect fluid. Indeed, the energy-momentum tensor is of the perfect fluid type, and the static energy functional is invariant under the base space SDiff transformations. 

A natural extension of the Skyrme model is its coupling to the electromagnetic field through the introduction of a $U(1)$ gauge field $A_\mu$ \cite{ED-sk}. Electromagnetic properties of nuclei form an essential part of nuclear physics, so this gauged version of the Skyrme model has obvious phenomenological applications, for recent results see, e.g., \cite{mar-man}. 

The baby Skyrme model, too, may be coupled to a $U(1)$ gauge field, and the standard procedure to introduce this coupling is straightforward. One just gauges the unbroken $O(2)$ subgroup by promoting the partial derivatives of the Skyrme field to covariant ones, $\partial_\mu \vec \phi \rightarrow D_\mu \vec \phi=(\partial_\mu +A_\mu \vec{n}\times )\vec \phi $ \cite{sch}. Obviously, this can be done both for the full baby Skyrme model and for the BPS submodel.
In the case of the BPS baby Skyrme model, however, there exist further natural possibilities to gauge it, and it is
the main aim of the present work to analyse the properties of these gauged versions of the BPS baby Skyrme model. Concretely, the gauging is performed by promoting the topological current (which is the building block in the derivative part of the action) to its gauged counterpart. This current, originally introduced by Schroers \cite{sch}, is uniquely defined by three conditions: gauge  invariance, conservation, and the condition that the space integral of its zero component (the gauged topological charge density) provides the correct topological charge. Quite surprisingly, this way of gauging the BPS baby Skyrme model leads to baby Skyrmions with qualitatively new properties. 

\section{Some known results}
The standard way of gauging the BPS baby Skyrme model leads to the following gauged BPS baby Skyrme model  \cite{gBPS} (coupling constants included)
\be
\mathcal{L}_{gauge \; baby \;BPS}= -\frac{\lambda^2}{4} \left(  \epsilon_{\mu \nu \rho} \left( \vec{\phi} \cdot D^\nu \vec{\phi} \times D^\rho \vec{\phi} \right)   \right)^2 - \mu^2V(\phi_3) - \frac{1}{4g^2} F_{\mu \nu}^2 .
\ee
It has been shown that the model keeps the BPS and perfect fluid properties. There is a topological bound
 \be
 E\geq 4\pi |n| \lambda^2 \left\langle W' \right\rangle_{\mathbb{S}^2}
 \ee
 which is saturated (therefore the theory is of the BPS type) for solutions of the following Bogomolnyi equations (see \cite{gBPS} for details)
 \bea
 \vec{\phi} \cdot D_1 \vec{\phi} \times D_2 \vec{\phi} &=& W' \\
B &=& -g^2\lambda^2 W.
\eea
Here, we have introduced an auxiliary target space function (the "superpotential") $W=W(\phi_3)$ defined by the so-called superpotential equation (a nonlinear differential equation in the target space coordinate $\phi_3$) 
\be
\lambda^2 W'^2 + g^2\lambda^4 W^2=2\mu^2 V.
\ee
Further, $B=F_{12}$ is the magnetic field and $n$ is the degree of the solution (topological charge). Moreover, the prime denotes differentiation with respect to $\phi_3$ and  $\left\langle W' \right\rangle_{\mathbb{S}^2}=(1/4\pi) \int d\Omega_{\mathbb{S}^2} W'$ is the target space average of $W'$. Let us briefly comment on the properties of the corresponding solutions (we refer to the original work \cite{gBPS}-\cite{bBPS-3}). 
First of all, such gauged BPS baby Skyrmions only exist if the superpotential equation has a solution defined on the full target space i.e., for $\phi_3 \in [-1,1]$. Due to the nonlinearity of the equation, this is a highly nontrivial question. For example, there are no solutions for double vacuum potentials, although non-gauged BPS baby Skyrmions do exist for these potentials. 
Secondly,  if skyrmions exist at all, they exist for any value of the topological charge $n$ and carry magnetic flux, $\Phi=\int d^2x B$, of a fixed value (for a given $n$). Specifically, the flux is linear in the topological charge, $\Phi=\Phi_0 n$. However, the proportionality constant $\Phi_0$ is never equal to $2\pi$. Therefore, such gauged solitons cannot be put on a compact manifold without boundary (a Riemannian surface). 

Solitons in the the gauged $O(3)$ sigma model, on the other hand, reveal rather different properties \cite{sch}. 
First of all, they are of BPS nature (solving a Bogomolnyi equation and saturating the corresponding bound) only for a unique potential $V=(1-\phi_3)^2$. Furthermore, these gauged BPS solitons only exist for topological charge greater that one, and they do not carry magnetic flux of a fixed value. Instead, there exists a one-parameter family of BPS solutions with a one-parameter family of fluxes, which never reach the value $2\pi n$. This, again, impedes to consider them on a compact manifold. 

In the general case, i.e., for the gauged version of the full baby Skyrme model (which, obviously, is not a BPS theory) one finds solitonic solutions with a fixed magnetic flux, where the flux is typically a nonlinear function of the topological charge \cite{GPS}, \cite{Shnir}. All this shows that, up to now, no gauged baby Skyrmions with a quantised magnetic flux $\Phi = 2\pi n$ have been found.  

\section{Schroers current}

The gauge invariant and conserved topological current relevant for baby Skyrmions reads \cite{sch}
\be
j_\mu = \frac{1}{2} \epsilon_{\mu \nu \rho} \left( \vec{\phi} \cdot D^\nu \vec{\phi} \times D^\rho \vec{\phi} + F^{\nu \rho} (1-\phi_3)\right)
\ee
where the covariant derivative is
$$D_\mu \vec{\phi} =  \partial_\mu \vec{\phi} +A_\mu \vec{n} \times \vec{\phi}$$
and 
$F_{\mu \nu} = \partial_\mu A_\nu - \partial_\nu A_\mu$. Note that this current is uniquely defined if an additional condition is imposed. Namely, that the degree of the map $\vec{\phi}$ is equal to 
\be
\mbox{deg } [\phi] =  \frac{1}{4\pi} \int d^2 x j_0
\ee
It is convenient to introduce the notation of \cite{gBPS},
\be
Q_\mu =  \frac{1}{2} \epsilon_{\mu \nu \rho}  \vec{\phi} \cdot D^\nu \vec{\phi} \times D^\rho \vec{\phi} 
\ee
\be
Q \equiv Q_0 =  \vec{\phi} \cdot D_1 \vec{\phi} \times D_2 \vec{\phi} 
\ee
and 
\be
j\equiv j_0= Q+F_{12}(1-\phi_3) = Q+B(1-\phi_3)
\ee
where $B=F_{12}$ is the magnetic field. For later use we will also define a deformed Schroers current 
\be
\tilde{j}_\mu = \frac{1}{2} \epsilon_{\mu \nu \rho} \left( \vec{\phi} \cdot D^\nu \vec{\phi} \times D^\rho \vec{\phi} + F^{\nu \rho} \sigma (\phi_3) \right)
\ee
where $\sigma (\phi_3)$ is at the moment an arbitrary function of the third component of the baby Skyrme field. This current is gauge invariant but, in general, not conserved, i.e., $\partial_\mu \tilde{j}^\mu \neq 0$. 
\section{The first model - the Schroers current squared}
\subsection{Topological bound and Bogomolny equations}
We will consider the following model
\be
\mathcal{L}= -\frac{\lambda^2}{2} j_\mu^2 - \mu^2 V(\phi_3) \label{model}
\ee
which is a natural gauged generalisation of the BPS baby Skyrme model, where the topological current is promoted to the Schroers current. 

The energy functional for the static, purely magnetic configurations is (we introduce an energy scale $E_0$)
\bea
E&=&\frac{1}{2} E_0 \int d^2x \left[ \lambda^2 j^2+2\mu^2 V(\phi_3)\right] =  \\
&=& \frac{1}{2} E_0 \int d^2x \left[ \lambda^2 Q^2 + \lambda^2 (1-\phi_3)^2 B^2+ 2\lambda^2 QB(1-\phi_3) +2\mu ^2 V \right]  \\
&=& \frac{1}{2} E_0 \int d^2x \left[ \lambda^2 Q^2 + \lambda^2 B^2 +2\mu ^2 V \right]  + \frac{\lambda^2 }{2} E_0 \int d^2x \left[ B^2\phi_3(\phi_3-2) + 2 QB(1-\phi_3)\right] .
\eea
We did not add the Maxwell term for the gauge field explicitly, but a Maxwell-type term is, nevertheless, induced. Indeed, the model can be viewed as the usual gauged BPS Skyrme model, with a particular relation between the coupling constants, and with two new coupling terms included: a "dielectric" like term where the Maxwell term (magnetic field squared) is coupled non-minimally to the matter field (we remark that such a modification of the gauged BPS baby Skyrme model has been very recently analysed in \cite{stepien-2}), and a new term which couples the topological current with the magnetic field. In the case at hand, the Maxwell-type term vanishes in the vacuum (for $\phi_3 \to 1$), but later we shall consider generalisations where this is not the case.

The BPS bound can be found in a similar fashion as in \cite{gBPS}. Consider the non-negative expression
\bea
0 \leq \frac{\lambda^2}{2}E_0 \int d^2 x \left[ \left( Q - w(\phi_3)\right)^2 + \left( (1-\phi_3)B +b(\phi_3)\right)^2 \right] &=& \\
\frac{\lambda^2}{2}E_0 \int d^2 x \left[  Q^2 +B^2(1-\phi_3)^2 +w^2 + b^2 - 2Qw + 2(1-\phi_3) Bb\right] &=& \\
\frac{\lambda^2}{2}E_0 \int d^2 x \left[  Q^2 +B^2(1-\phi_3)^2 +w^2 + b^2 -2qw \right]  &+&\\
+ \lambda^2 E_0 \int d^2 x \left[  - w \epsilon_{ij}A_i \partial_j \phi_3 + (1-\phi_3) b \epsilon_{ij} \partial_i A_j \right]
\eea
The last line can be written as a total derivative if
\be
(1-\phi_3) b(\phi_3) = W=\int_{\phi_3=1}^{\phi_3} dt \; w(t) \;\;\; \Rightarrow \;\;\; W'=w
\ee
where we assume that the vacuum value of the baby Skyrme field is $\phi_3=1$. Furthermore, we used that
\be
Q=q+\epsilon_{ij}A_i \partial_j \phi_3, \;\;\; q=\vec{\phi} \cdot \partial_1 \vec{\phi} \times \partial_2 \vec{\phi} .
\ee
Indeed, 
\bea
- w \epsilon_{ij}A_i \partial_j \phi_3 + (1-\phi_3) b \epsilon_{ij} \partial_i A_j &=&
- W' \epsilon_{ij}A_i \partial_j \phi_3 + W \epsilon_{ij} \partial_i A_j = \nonumber \\
- (\epsilon_{ij} A_i \partial_j W -\epsilon_{ij} W \partial_i A_j ) &=&
- (\epsilon_{ij} A_i \partial_j W +\epsilon_{ij} W \partial_jA_i) \; =\;  - \epsilon_{ij} \partial_j (A_i W) .
\eea
Then,
\be
\int d^2 x \epsilon_{ij} \partial_j (A_i W) =0 
\ee
as $W(\phi_3=1)=0$ by construction. Hence,
\be
\frac{\lambda^2}{2}E_0 \int d^2 x \left[  Q^2 +B^2(1-\phi_3)^2 +\frac{W^2}{(1-\phi_3)^2} + W'^2 \right] \geq \lambda^2 E_0 \int d d^2 x qw .
\ee
The left hand side is our energy functional if we impose that 
\be
\lambda^2 W'^2 +\lambda^2 \frac{W^2}{(1-\phi_3)^2} = 2\mu^2 V + 2\lambda^2 QB(1-\phi_3) .
\ee
This is a version of the previously found superpotential equation. In its current form, however, it is not yet a pure target space ($\mathbb{S}^2$) equation since it mixes the baby Skyrme (matter) field with the gauge field. 
Finally we get the following bound
\be
E \geq \lambda^2 E_0 \int d^2 x qW' = 4\pi |n| E_0\lambda^2 \left\langle W' \right\rangle_{\mathbb{S}^2} .
\ee
The bound is saturated if the following Bogomolnyi equations are satisfied 
\bea
Q &=& W' \\
(1-\phi_3) B &=& -\frac{W}{1-\phi_3} . \label{Bog}
\eea
Now, the superpotential equation can finally be rewritten as an on-shell $\mathbb{S}^2$ target space equation. Indeed, using the Bogomolnyi equations we may eliminate $Q$ and $B$ and find
\bea
\lambda^2 W'^2 +\lambda^2 \frac{W^2}{(1-\phi_3)^2} - 2\lambda^2 QB(1-\phi_3) &= &2\mu^2 V \nonumber \\
\lambda^2 \left( W'^2 +  \frac{W^2}{(1-\phi_3)^2}  + \frac{2WW'}{1-\phi_3}  \right) &=& 2\mu^2 V \nonumber \\
\lambda^2 \left( W' + \frac{W}{1-\phi_3}  \right)^2 &=& 2\mu^2 V .
\eea
Hence, it can be reduced to a first order {\it linear } differential equation 
\be
W' + \frac{W}{1-\phi_3} =\pm \frac{\mu}{\lambda} \sqrt{2V(\phi_3)} .
\ee
This should be contrasted with the former version of the gauged BPS baby Skyrme model, where the superpotential equation has a highly 
nonlinear form, making its analytical treatment very complicated. 
The corresponding general solution is a sum of the general solution of the homogenous part and a particular solution 
of the inhomogenous one.
The homogenous part 
\be
W' + \frac{W}{1-\phi_3} =0
\ee
has a simple solution 
\be
W= A(1-\phi_3)
\ee
where $A$ is an arbitrary constant. Note that all these solutions obey the boundary condition $W(\phi_3=1)=0$. Nonetheless, we will show in an example that only one solution of the super-potential equation (with the boundary condition) is physically acceptable.

As an example, we will consider the following family of potentials $V=\frac{1}{2} (1-\phi_3)^{2k}$. Then, the super-potential obeying $W(1)=0$ is
\be
W=A(1-\phi_3) \mp \frac{1}{k} \frac{\mu}{\lambda}(1-\phi_3)^{k+1} .
\ee
It is straightforward to notice that the solutions with $A \neq 0$ are not acceptable since they lead to singular magnetic fields when the baby Skyrme field takes the vacuum value. 
\subsection{$\kappa$ generalisation}
A simple but interesting observation is that the Schroers current is not uniquely defined if we we impose only two out of three properties: gauge invariance and conservation. Indeed, one can add a purely gauge current with any coupling constant $\kappa$ without spoiling these properties
\be
j_\mu \rightarrow j_\mu +\frac{\kappa -1}{2} \epsilon_{\mu \nu \rho} F^{\nu \rho}.
\ee
On the other hand, the integrated charge density of this current no longer provides the topological charge (it receives contributions from the magnetic flux). 
However, in our construction of BPS models,  this fact is not problematic.
Hence, the most general current reads 
\be
j_\mu = \frac{1}{2} \epsilon_{\mu \nu \rho} \left( \vec{\phi} \cdot D^\nu \vec{\phi} \times D^\rho \vec{\phi} + F^{\nu \rho} (\kappa-\phi_3) \right) .
\ee
The nice feature is that for $\kappa >1$ the vacuum value of the baby Skyrme field (assumed as $\phi_3=1$) does not kill the second part of the current identically but leaves the free gauge part. (Obviously, it also dynamically disappears as the magnetic field tends to 0). This means, at the level of the action, that assuming the vacuum of the matter field we find the free Maxwell Lagrangian. Therefore, asymptotically, we deal with the usual (not dielectrically deformed) electromagnetic field.  
\\
For the model based on this generalised current we get the same bound which is saturated if (we restrict to the case when $\kappa >1$)
 \bea
Q &=& W' \\
(\kappa-\phi_3) B &=& -\frac{W}{\kappa-\phi_3}
\eea
where
\be
W' + \frac{W}{\kappa-\phi_3} =\pm \frac{\mu}{\lambda} \sqrt{2V(\phi_3)} .
\ee
For example, for the following potential $V=\frac{1}{2} (1-\phi_3)^2$ the solution is
\be
W=\mp  \frac{\mu}{\lambda} (\kappa - \phi_3) \left[ (1-\kappa) \ln \left( \frac{\kappa - \phi_3}{\kappa-1} \right) +1-\phi_3 \right] .
\ee
Here the boundary condition is enough to uniquely fix the solution. No additional (finite magnetic flux) condition is needed. 
\subsection{The magnetic flux}
To compute the magnetic flux, for simplicity we assume the axial symmetry (although, owing to the symmetries of the model, the final result holds for arbitrary solutions) 
\be
\vec{\phi} (r,\varphi)  = \left(
\begin{array}{c}
\sin f(r) \cos n \varphi \\
\sin f(r) \sin n \varphi \\
\cos f(r) 
\end{array}
\right), \;\;\; A_0=A_r=0, \;\;\; A_\varphi=na(r) .
\ee
We also introduce the new base space 
\be
y=\frac{1}{2}r^2
\ee
and target space coordinate
\be
h = \frac{1}{2} (1-\phi_3) .
\ee
Then, the Bogomolnyi equations for $\kappa=1$ model (\ref{Bog}) take the following form 
\bea
2n h_y(1+a)&=&-\frac{1}{2} W_h \\
2h na_y &=& -\frac{W}{2h} .
\eea
If we divide the first equations by the second we find
\be
\frac{h_y}{h} \frac{(1+a)}{a_y} = \frac{W_h h}{W} \;\;\; \Rightarrow \;\;\; \frac{h_y W}{W_h h^2} = \frac{a_y}{(1+a)} .
\ee
This can be integrated as
\be
\ln C(1+a) = F(h), \;\;\;\;\; F(h) \equiv \int_0^h dh' \frac{W}{W_{h'} h'^2} .
\ee
Taking into account the boundary conditions $a(y=0)=0$ and $h(y=0)=1$ we get
\be
C=e^{F(h=1)}
\ee
and
\be
a = -1 +e^{-F(1)+F(h(y))} .
\ee
Now, we can compute the magnetic flux
\be
\Phi= \int rdr d\varphi B = 2\pi n \int dy a_y = 2\pi n a_\infty
\ee
where $a_\infty$ is the value at the infinity or at the compacton boundary, where $h=0$. Hence,
\be
a_\infty = -1+e^{-F(1)+F(0)}=-1+e^{-F(1)}.
\ee
One should underline that, in opposition to the gauged $O(3)$ model, the magnetic flux is a fixed function linear in the topological charge, $\Phi=\Phi_0 n$, since $a_\infty$ takes one unique value determined by the solutions of the superpotential equation. This property is shared also by the previously considered gauged BPS baby Skyrme models. However, $\Phi/2\pi$ was not an integer number, as $F(1)$ takes a finite value. Therefore, such a theory cannot be defined on a compact manifold. This feature is no longer true for our new class of models (\ref{model}). 

Let us consider a particular example and turn to the formerly considered family of potentials
\be
V=  \frac{1}{2} (2h)^k
\ee
which lead to the following superpotential
\be
W=\pm \frac{1}{k} \frac{\mu}{\lambda} (2h)^{k+1}.
\ee
Then,
\be
-F(1)\equiv- \int_0^1 dh \frac{W}{W_{h} h^2} = - \frac{1}{k+1} \int_0^1 \frac{dh}{h} = -\infty .
\ee
This results in $a_\infty = -1$. This results in the surprising fact that this model supports gauged baby Skyrmion solutions with quantised magnetic flux, i.e., $\Phi/ 2\pi$ is an {\it integer} number. Note that this is the first baby Skyrme type model with this property. It follows that these gauged baby Skyrmions can be considered on compact manifolds. 

Observe also that the $\kappa$ deformed model with $\kappa >1$ does not possess this property. The corresponding expression for the flux is 
\be
a_\infty = -1 +e^{-F(1)}, \;\;\; -F(1)=- \int_0^1 dh \frac{W}{W_{h} (\kappa-1+2h)^2} .
\ee 
But now, since $\kappa-1+2h >0$, the integral is finite. 
\subsection{Solutions}
Let us now solve exactly the $\kappa=1$ model with the potentials introduced above, for the axially symmetric ansatz. Specifically, 
\be
F(h)=\frac{1}{k+1} \ln h \;\;\; \Rightarrow \;\;\; a=-1 + h^{\frac{1}{k+1}}
\ee
or
\be
h=(1+a)^{k+1} .
\ee
Then, inserting this into the second Bogomolnyi equation, we get the following solution: for $k< \sqrt{2}$ we find 
compactons
\be
a=-1 + \left( 1-\frac{y}{y_R}\right)^{\frac{1}{2-k^2}}, \;\;\;\; y_R=n \frac{\lambda}{\mu} \frac{k}{2^{k-1} (2-k^2)}
\ee
where $y_R$ is the compacton radius, while for $k>\sqrt{2}$ power-like localised solitons
\be
a=-1 + \left( \frac{y_1}{y_1+y} \right)^{\frac{1}{k^2-2}}, \;\;\;\; y_1=n \frac{\lambda}{\mu} \frac{k}{2^{k-1} (k^2-2)} .
\ee
For $k = \sqrt{2}$ we get an exponential solution
\be
a=-1+e^{-\frac{2^{\sqrt{2}-1}}{\sqrt{2}} \frac{\mu}{n \lambda } y} .
\ee
Using the algebraic relation between $a$ and $h$ we can easily find the profile of the baby Skyrmions. 
\subsection{Double vacuum potential}
It was a surprising fact for the original gauged BPS baby Skyrme model \cite{gBPS}  that there are no
soliton solutions for a double-vacuum potential, although they exist for the non-gauged model. 
Here we will show that this is a rather model dependent result which is no longer true for a gauged BPS baby Skyrme type model constructed by means of the Schroers current (\ref{model}). 
\\
In the following example we consider the potential
\be
V=\frac{1}{2} (1-\phi_3)^2(1+\phi_3)^2
\ee
with two vacua $\phi_3= \pm 1$. Then the superpotential equation 
\be
W'+\frac{W}{1-\phi_3}=\pm \frac{\mu}{\lambda}(1-\phi_3)(1+\phi_3)
\ee
has a unique finite flux solution
\be
W=\mp \frac{1}{2}\frac{\mu}{\lambda} (3+\phi_3)(1-\phi_3)^2
\ee
existing on the whole interval $\phi_3 \in [-1,1]$. Thus, the integral 
\be
-\int_0^1dh \frac{W}{W_h h^2} = -4\int_{-1}^1 d\phi_3 \frac{(3+\phi_3)}{(5+3\phi_3)(1-\phi_3)}=-\infty
\ee
which results in the quantised magnetic flux $\Phi/2\pi \in \mathbb{Z}$. Now, we can solve the Bogomolnyi equations in the axial ansatz
\bea
2n h_y(1+a)&=&-\frac{1}{2} W_h \\
2h na_y &=& -\frac{W}{2h}, \;\;\; W=4h^2(2-h) .
\eea
Following the previous strategy we first find an algebraic relation between the magnetic and skyrmionic functions
\be
a=-1+e^{-\int_h^1 dh \frac{W}{W_h h^2}} = -1 +e^{\frac{1}{6} \ln \frac{h^3}{4-3h}}=-1+ \frac{h^{1/2}}{(4-3h)^{1/6}} .
\ee
Then, the solution for the Skyrmion profile function is given by an integral
\be
\int_0^h \frac{dh'}{h'^{1/2}(4-3h')^{7/6}} = \frac{\mu}{n \lambda } (y_R-y) 
\ee
where the radius of the compacton $y_R$ follows from the condition $h(y=0)=1$
\be
\int_0^1 \frac{dh'}{h'^{1/2}(4-3h')^{7/6}} = \frac{\mu}{n \lambda } y_R .
\ee
Numerically it gives
\be
y_R\approx 0.6678\frac{n\lambda}{\mu} .
\ee
\section{The second model - the deformed current squared}
All results can be repeated for the deformed current model
\be
\mathcal{L}= -\frac{\lambda^2}{2} \tilde{j}_\mu^2 - \mu^2 V(\phi_3) .
\ee
Here the static energy is
\be
E=  \frac{1}{2} E_0 \int d^2x \left[ \lambda^2 Q^2 + \lambda^2 \sigma^2 B^2+ 2\lambda^2 \sigma QB  +2\mu ^2 V \right] .
\ee
The bound has exactly the same form 
\be
E \geq  \lambda^2 E_0 \int d^2 x qW' = 4\pi |n| E_0\lambda^2 \left\langle W' \right\rangle_{\mathbb{S}^2}
\ee
but the function $W$ is defined by a different superpotential equation containing the deformation function $\sigma$,
\be
\lambda^2 W'^2 +\lambda^2 \frac{W^2}{\sigma^2} = 2\mu^2 V + 2\lambda^2 \sigma QB .
\ee
The bound is saturated for solutions of the Bogomolnyi equations
\bea
Q &=& W' \\
\sigma B &=& -\frac{W}{\sigma}
\eea
Then, the on-shell superpotential equation reads
\be
W' + \frac{W}{\sigma(\phi_3)} = \frac{\mu}{\lambda} \sqrt{2V(\phi_3)}, \;\;\;\;\; W(\phi=1)=0.
\ee
The uniqueness of the solution is related to the form of $\sigma$ and, in the case when $\sigma$ has a zero at $\phi_3=1$, the condition of  finiteness of the magnetic flux must be added. On the other hand, the existence of a solution on the whole segment $\phi_3 \in [-1,1]$ is a mutual effect of $\sigma$ and $V$. However, for reasonable deformations and potentials such a $W$ exists. 
\section{The most general SDIff invariant gauged baby BPS Skyrme model}
The most general Lagrangian density which can support gauged baby Skyrmions obeying two requirements:
1) describes a perfect fluid matter (i.e., the energy-momentum tensor has the form of a perfect fluid and  the static, purely magnetic energy functional is (base space) SDiff invariant; 2) there is a topological bound which is saturated by solutions of the pertinent Bogomolny equations, reads
\be
\mathcal{L}= \frac{1}{2} g_1 Q_\mu^2 + \frac{1}{4}g_2 \epsilon^{\mu \nu \rho} Q_\mu F_{\nu \rho} +\frac{1}{4} g_3 F_{\mu \nu}^2 + \mu^2V
\ee
where $g_1,g_2,g_3$ are coupling functions depending on $\phi_3$ (and containing coupling constants). The corresponding static energy 
is
\be
E=\frac{E_0}{2} \int d^2 x \left[ g_1 Q^2 + g_2 QB + g_3 B^2 +2\mu^2  V \right] .
\ee
In order to derive a topological bound we follow the same path as in the previous cases. Consider th following non-negative expression
\bea
0\leq \frac{E_0}{2}\int d^2x  \left[ \left(\sqrt{g_1} Q - w(\phi_3) \right)^2 + \left(\sqrt{g_3} B +b(\phi_3) \right)^2 \right] &=& \\
\frac{E_0}{2}\int d^2x  \left[ g_1 Q^2 + g_3 B^2 + w^2 +b^2 - 2q w\sqrt{g_1} \right] &+& \\
E_0 \int d^2 x \left[ -w\sqrt{g_1} \epsilon_{ij} A_i \partial_j \phi_3 + b\sqrt{g_3} \epsilon_{ij} \partial_i A_j\right] .
\eea
The last line vanishes as the integrand is a total derivative if
\be
b\sqrt{g_3} = W = \int_{\phi_3=1}^{\phi_3} dt \; w(t) \sqrt{g_1(t)} \;\;\;\; \Rightarrow \;\;\;\; W'=w\sqrt{g_1} .
\ee
Therefore,
\be
\frac{E_0}{2} \int d^2 x  \left[ g_1 Q^2 + g_3 B^2 + \frac{W'^2}{g_1} +\frac{W^2}{g_3} \right] \geq  E_0 \int d^2 x \; q w\sqrt{g_1} .
\ee
Assuming that $W$ obeys 
\be
\frac{W'^2}{g_1} +\frac{W^2}{g_3} =  2\mu^2 V + g_2 QB
\ee
we get
\be
E\geq \int d^2 x W' = 4\pi |n|  \left\langle W' \right\rangle_{\mathbb{S}^2} .
\ee
Obviously the bound is saturated if the fields obey the following Bogomolnyi equations 
\bea
\sqrt{g_1}Q &=& \frac{W'}{\sqrt{g_1}} \\
\sqrt{g_3} B &=& - \frac{W}{\sqrt{g_3}} .
\eea
Now, in the Bogomolnyi sector,  we may simplify the superpotential equation to a form which involves only the Skyrme field (target space equation)
\be
\frac{W'^2}{g_1} +\frac{W^2}{g_3} +g_2 \frac{W'W}{g_1g_3}=  2\mu^2 V .
\ee

\section{Deformed gauged $O(3)$ model}
Finally, we want to comment on the $O(3)$ gauged model with a potential which is known to 
support BPS solitons. Here we very briefly summarise the results of \cite{sch}. Let us consider only the static purely magnetic energy integral
\be
E=\frac{1}{2}E_0 \int d^2 x \left[ (D_1 \vec{\phi})^2+(D_2 \vec{\phi})^2 +(1-\phi_3)^2 +B^2 \right] .
\ee 
The corresponding bound and Bogomolny equation have been obtained in \cite{sch}. The proof is as follows. The energy can be written as
\be
E=\frac{1}{2}E_0 \int d^2 x \left[ (D_1 \vec{\phi} \pm \vec{\phi} \times D_2 \vec{\phi})^2 + (B\pm (1-\phi_3))^2 \right] \pm E_0 \int d^2 x (\vec{\phi} \cdot D_1\vec{\phi} \times D_2\vec{\phi} + B (1-\phi_3) )
\ee
where the last integral contains the Schroers density. Hence,
\be
E\geq 4\pi E_0 |n|
\ee
where the equality occurs for solutions of the Bogomolnyi equations
\bea
D_1 \vec{\phi} &=& \mp \vec{\phi} \times D_2 \vec{\phi} \\
B&=&\pm (1-\phi_3)
\eea
which after the stereographic projection read
\bea
D_1 u \pm i D_2 u&=&0 \\
B&=& \pm \frac{2|u|^2}{1+|u|^2} .
\eea
It was shown that the gauged $O(3)$ lumps exist for $|n| \geq 2$ and the magnetic flux is not quantised. In fact, the flux has no fixed value as $a_\infty \in [-1,0)$ \cite{sch}. Instead, we have a one-parameter family of solutions with a one-parameter family of fluxes. 

It is straightforward to notice that this construction can be generalised to the following deformed model
\be
E=\frac{1}{2}E_0 \int d^2 x \left[ (D_1 \vec{\phi})^2+(D_2 \vec{\phi})^2 +\frac{1}{g^2}(1-\phi_3)^2 +g^2 B^2 \right]
\ee 
where $g=g(\phi_3)$ is an arbitrary function. Then, repeating the computations above, we find that again
\be
E \geq 4\pi E_0 |n| .
\ee
The bound is saturated if
\bea
D_1 \vec{\phi} &=& \mp \vec{\phi} \times D_2 \vec{\phi} \\
gB&=&\pm \frac{1}{g} (1-\phi_3)
\eea
or in the complex $u$ field version
\bea
D_1 u \pm i D_2 u&=&0 \\
g(|u|) B&=& \pm\frac{1}{g(|u|)}  \frac{2|u|^2}{1+|u|^2} .
\eea
One can immediately ask whether properties of the BPS gauged solitons in the undeformed model ($g=1$) persist in the deformed version (nontrivial $g$). 

First of all, let us observe that the deformation changes only the second Bogomolny equation, while the first one remains the same. Therefore, there is still a one parameter family of solutions. To show this we assume the axial ansatz. Then, the equations are (here $\phi_3 = \cos f$)
\bea
f' &=&-|n| \frac{1+a}{r} \sin f \\
g(f) a' &=& - \frac{r}{|n|} \frac{(1-\cos f)}{g(f)}.
\eea
Assuming the standard boundary conditions at the origin, $f(r=0)=\pi$, $a(r=0)=0$ we get that
$f\approx\pi +Ar^{|n|}$. Hence, there is still a one-parameter family of solutions of the Bogomalny equations (with the boundary conditions). This is reflected in the existence of topological solutions with a continuous value of the magnetic flux. In other words, the magnetic flux is still not fixed and, therefore, it is not quantised. 

In the next step, we consider a particular deformation provided by $g=(1-\cos f)^{\beta}$. Then, asymptotically at infinity (or at the soliton boundary in the case of compactons), we get 
\be
f'=-|n| \frac{1+a_\infty}{r}f, \;\;\;\; a'=-\frac{r}{|n|} \left( \frac{f^2}{2} \right)^{1-2\beta} .
\ee
Hence,
\be
a' \sim r^{1-2|n|(1-2\beta)(1+a_\infty)}
\ee
which gives a non-singular $a$ if
\be
(1-2\beta) (1+a_\infty) > \frac{1}{|n|} .
 \ee
Following the proof in \cite{sch}, one can again show that $a$ is a monotonously decreasing function with $a_\infty \in [-1, 0)$. Assuming $n=1$ we get a condition for $\beta$
\be
\beta < \frac{a_\infty}{2(1+a_\infty)}
\ee
Therefore, for sufficiently small $\beta$ charge one solitons exist. 

Let us remark that formally equivalent Bogomolnyi equations have been recently found in \cite{guilarte}, however, in the context of models supporting vortices, i.e., with a different topology. This means that the potential has a one-dimensional vacuum manifold, defined by $|u|=v=\mbox{ const.}$, which is equivalent to a circle $\mathbb{S}^1 \subset \mathbb{C}$ (or $\mathbb{S}^1 \subset \mathbb{S}^2$ if we consider the original $\vec{\phi}$ vector field). As a consequence, the "Higgs" field covers only the fundamental domain in the complex space, $|u| \leq v$, and the magnetic flux is necessarily quantised in $2\pi$ units as it happens, for example, for the Abelian Higgs model. See also \cite{bazeia}.
\section{Summary}
In the present work, we have studied a new way of gauging the BPS baby Skyrme model, by promoting the topological current to its gauged version, the so-called Schroers current. This is the unique current obeying the three conditions that {\em i)} it is gauge invariant, {\em ii)} it is conserved and {\em iii)}  its charge gives the correct degree of the map. The resulting Lagrangian differs in several aspects from the gauged BPS baby Skyrme model introduced previously in \cite{gBPS}. For example, it contains two new terms: a non-minimal, dielectric-like coupling between the matter and the gauge field and a new interaction between the magnetic field and the topological density. Of course, all terms in the action enter with very specific couplings, which allows to combine them into the square of the Schroers current, plus a potential. 

This new gauged BPS baby Skyrme model possesses very distinctive features if compared with the old gauged BPS baby Skyrme model (or other gauged baby Skyrme theories). 
First of all, the model is completely solvable for any value of the topological charge. This follows from a special property of the superpotential equation which supplements the Bogomolny equations. Namely, in contrast to former models, it is a {\it linear} differential equation in the target space variable $\phi_3$ which can easily be solved for any potential. As a consequence, gauged solitons exist for any reasonable potential, also with two vacua. This is different from the (surprising) finding in the old version of the gauged BPS baby Skyrme model, where solitons in two-vacuum potentials were forbidden. We see that this fact is related to the specific form of the action (reflected in the superpotential equation) rather than anchored in some deeper mathematical obstacles of gauged BPS baby Skyrme type theories. 

Secondly, the baby skyrmions found in the model (\ref{model}) carry quantised magnetic flux, i.e., $\Phi=2\pi n$. This originates from the fact that the superpotential equation has a singularity at the vacuum $\phi_3=1$. As a consequence, this model can be considered on a compact manifold without boundary. We want to underline that this is the unique baby Skyrme type model (both BPS or not BPS) with such a property. 


Moreover, we found Bogomolnyi equations and the corresponding superpotential equation for some generalised or deformed models, where the only requirement for them (besides the fact that they should possess a BPS sector) is the perfect fluid property. Finally, we analysed a deformed gauged $O(3)$ model, showing that the magnetic flux remains non-quantised (and does not even assume a fixed value). 

We remark that a gauge-invariant and conserved version of the topological current (analogously to the Schroers current considered here) may be defined for Skyrmions in 3+1 dimensions in the presence of an abelian gauge field (the so-called Goldstone-Wilczek current \cite{GoldWil}). Lagrangians based on this current may, therefore, also be defined for Skyrmions in 3+1 dimensions and offer new possibilities to gauge Skyrmions in 3+1 dimensions. This issue is, however, beyond the scope of the present paper. 

\section*{Acknowledgements}
The authors acknowledge financial support from the Ministry of Education, Culture, and Sports, Spain (Grant No. FPA 2014-58-293-C2-1-P), the Xunta de Galicia (Grant No. INCITE09.296.035PR and Conselleria de Educacion), the Spanish Consolider-Ingenio 2010 Programme CPAN (CSD2007-00042), and FEDER.  AW was supported by NCN grant 2012/06/A/ST2/00396. We thank Wojtek Zakrzewski for comments and for reading the manuscript.

\end{document}